\documentclass[12pt]{article}
 \usepackage{graphicx}
 \usepackage[cp1251]{inputenc}
 \usepackage{rotating}
\begin{document}
 \noindent {\footnotesize\it Astronomy Letters, 2025, Vol. 51, No. 8}
 \newcommand{\dif}{\textrm{d}}

 \noindent
 \begin{tabular}{llllllllllllllllllllllllllllllllllllllllllllll}
 & & & & & & & & & & & & & & & & & & & & & & & & & & & & & & & & & & & & & &\\\hline\hline
 \end{tabular}

 \vskip 0.2cm
 \bigskip
  \centerline{\bf\Large  A new estimate of the zero-point shift of the Gaia DR3 }
  \centerline{\bf\Large  parallaxes obtained from a comparison with VLBI }
  \centerline{\bf\Large measurements of masers and radio stars}
\bigskip
 \centerline{\bf   Vadim V. Bobylev\footnote[1]{bob-v-vzz@rambler.ru}   }
 \bigskip
 \centerline{\it Central (Pulkovo) Astronomical Observatory, Russian Academy of Sciences}
 \bigskip
 \bigskip
{The most complete sample of radio stars and masers with trigonometric parallaxes measured by the VLBI method, common with the Gaia\,EDR3 and Gaia\,DR3 catalogs, has been compiled using literature data. The sample contains 151 stars. An analysis of the differences in parallaxes and proper motions of Gaia--VLBI stars has been performed. A new estimate of the systematic shift of the Gaia parallax zero point relative to the inertial coordinate system has been obtained: $\Delta\pi=-0.038\pm0.011$~mas. The obtained estimates of the relative rotation rates confirm the absence of a significant rotation of the Gaia\,DR3 system relative to the extragalactic coordinate system, which in this case is represented by VLBI measurements.
  }

 \bigskip
 \section*{INTRODUCTION}
 The zero point of the Gaia (Gaia Collab. 2016) trigonometric parallaxes  has an offset of $\Delta\pi=-0.029$~mas (milliarcseconds) relative to the quasar system (Lindegren et al. 2018). This offset has been repeatedly confirmed based on a variety of data using different Gaia releases -- Gaia\,DR2 (Gaia Collab. 2018), Gaia\,EDR3 (Gaia Collab. 2021), and Gaia\,DR3 (Gaia Collab. 2023). Objects with an independent distance scale are used to confirm this offset, for example, variable stars, the distances to which are calculated based on the period-luminosity relation; binaries with known orbits, allowing us to estimate their dynamical (orbital) parallaxes; Masers and radio stars, the trigonometric parallaxes of which were measured by the VLBI method with reference to distant quasars; asteroseismic parallaxes of red giants; red clump stars, etc.

A comparison of classical Cepheids with the Gaia\,DR2 catalog was performed, for example, in the papers of Riess et al. (2018) and Groenewegen (2018). Using data on 46 Cepheids, Riess et al. (2018) found $\Delta\pi=-0.046\pm0.013$~mas. Using only nine Cepheids with very reliable distances, Groenewegen (2018) obtained the estimate $\Delta\pi=-0.049\pm0.018$~mas. Using a sample of approximately 400 RR\,Lyrae variables, the correction $\Delta\pi=-0.057\pm0.006$~mas was found in the paper by Muraveva et al. (2018). A comparison of the absolute magnitudes of 147 RR\,Lyrae variables by Leiden et al. (2019) yielded $\Delta\pi = -0.042\pm 0.013$~mas.

Ren et al. (2021) analyzed a sample of W Ursae Majoris variables from the Gaia EDR3 catalog. Using a refined period-luminosity relation, parallaxes were estimated for 1,194 such stars with errors averaging about 7.5\%. These authors ultimately found
$\Delta\pi=-0.029\pm0.001$~mas.

Based on 89 separated eclipsing binaries shared with the Gaia DR2 catalog, Stassum and Torres (2018) found a correction of $\Delta\pi = -0.082\pm 0.033$~mas. Based on a comparison of dynamical parallaxes obtained from an analysis of the orbits of 76 eclipsing binaries in the Gaia\,EDR3 catalog, Stassum and Torres (2021) estimated $\Delta\pi = -0.037\pm 0.020$~mas. These authors report that their dynamical parallaxes were determined with average errors of less than 15\%. To study the Gaia\,DR2 zero-point shift, Grachik et al. (2019) used a sample of 81 separated eclipsing binaries. They analyzed photometric parallaxes, rather than orbital ones, and found $\Delta\pi = -0.054\pm 0.024$~mas.

In the paper by Zinn et al. (2019), independent confirmation of the parallax zero-point bias of Gaia\,DR2 was presented, found using asteroseismic data on evolved stars in the field of the Kepler space telescope. For this, data on approximately 3000 red giants from the APOKASC-2 catalog (Apache Point Observatory Kepler Asteroseismology Science Consortium, Pinsonneault et al., 2018) were used, from which a correction of $\Delta\pi=-0.053\pm0.009$~mas was found. The same 3000 red giants from the Kepler field with the addition of 2200 red clump giants were also used in the paper by Han et al. (2019), where the estimate of Zinn et al. (2019) was essentially confirmed. The parallax zero-point correction value found by Zinn et al. (2021) using a similar method from a comparison with data from the Gaia EDR3 catalog was $\Delta\pi=-0.015\pm0.002$~mas.

By comparing the distances of approximately 150,000 stars from the Gaia\,DR2 catalog with their spectrophotometric distances obtained from data from the APOGEE (The Apache Point Observatory Galactic Evolution Experiment, Majewski et al. 2017) survey, Ljung and Bovy (2020) obtained an estimate of $\Delta\pi=-0.052\pm0.002$~mas. And for the stars belonging to the red giant clump on the Hertzsprung-Russell diagram, Chan and Bovy (2020) found $\Delta\pi=-0.048\pm0.001$~mas. From a comparison with the Gaia\,EDR3 catalog of about 60,000 red giants from the LAMOST (Large sky Area Multi-Object fiber Spectroscopic Telescope, Deng et al. 2012) spectroscopic survey, Huang et al. (2021) showed that the shift $\Delta\pi$ is about $-0.026$~mas, the value of which strongly depends on the stellar magnitude $G$, the shape of the spectrum (color) and the position of the stars on the celestial sphere. About 50,000 LAMOST giants were used in the paper by Wang et al. (2022), where $\Delta\pi=-0.028$~mas from a comparison with stars from the Gaia\,EDR3 catalog (for a five-parameter solution) was found. Here, the error $\Delta\pi$ is very small, it can be calculated as $\sigma/\sqrt n$, where $\sigma=0.057$~mas is the variance given by the authors in their Fig.~4, and $n=50,000$ is the number of stars in the sample.

Many of the optical methods noted above, in one way or another, rely on calibrations based on trigonometric parallaxes determined by optical methods. However, measurements are now available that implement an independent inertial coordinate system with high astrometric accuracy. These include measurements of maser sources and radio stars using Very Long Baseline Interferometry (VLBI) methods, performed with reference to quasars.

Bobylev (2019) found $\Delta\pi=-0.038\pm0.046$~mas for the ``Gaia\,DR2--VLBI'' differences by analyzing a sample of 75 masers and radio stars with VLBI measurements of their trigonometric parallaxes. However, it turned out that achieving more acceptable accuracy requires significantly reducing the sample size. As a result, based on 49 of the most reliable stars, whose parallax difference moduli do not exceed 0.25 mas, an estimate of $\Delta\pi=-0.035\pm0.025$~mas was obtained. Xu et al. (2019) repeated the analysis of this sample, used a slightly different weighting system, proposed discarding stars of the asymptotic giant branch, leaving only 33 young objects in the sample, and found $\Delta\pi=-0.075\pm0.029$~mas based on differences of the type ``Gaia\,DR2--VLBI''. Based on 41 radio stars with VLBI measurements of their trigonometric parallaxes, Lindegren (2020), having developed a fairly effective system for rejecting poor residuals, obtained an estimate of $\Delta\pi=-0.076\pm0.025$~mas using the ``Gaia\,DR2--VLBI'' type of residuals. By comparing with the Gaia\,EDR3 catalog, Bobylev (2022) found $\Delta\pi=-0.022\pm0.017$~mas for a sample of 90 masers and radio stars.

Recently, results have appeared from an analysis of masers and radio stars whose parallaxes were measured using the VLBI method, where $\Delta\pi$ estimates were obtained with very high accuracy. Thus, Andriantsaralaza et al. (2022) performed an analysis of 33 oxygen-rich stars belonging to the Asymtotic Giant Branch (AGB) and possessing maser emission. Using 17 such stars (with stellar magnitudes $G<8^m$), an estimate of $\Delta\pi=-0.077\pm0.007$~mas was obtained, and for all 33 stars of this sample, $\Delta\pi=-0.131^{+0.016}_{-0.015}$~mas was found using differences of the type ``Gaia\,DR2--VLBI''.

%%%%%%%%%%%%%%%%%%%%%%%%%%%%%%%%%%%%%%%%%%%%%
{
\begin{table}[t]
\caption[]{\small   Data on trigonometric parallaxes of masers and radio stars. }
\begin{center} \label{t-DATA-pi}
\begin{tabular}{|r|c|c|c|c|c|c|c|c|}\hline
              Star  &  $\alpha$, deg.  &  $\delta$, deg.  &   $\pi_{\rm VLBI}$, mas & $\pi_{\rm GAIA}$, mas &  $G$  & $B$ & $R$ \\\hline
SY Scl                 &  01.90104 &  $-25.4944$  & $ ~0.750\pm0.030$ & $ ~0.5247\pm0.1216$ & 10.0 & 3 & a \\
S Per                   &  35.71545 &  $+58.5865$ & $ ~0.413\pm0.017$ & $ ~0.2217\pm0.1214$ &  7.7 & 3 & b \\
HIP 12469            &  40.13193 &  $+61.2293$ &                      ---    & $ ~0.3776\pm0.0130$ & 10.3 & 2 & c \\
RZ Cas                &  42.23132 &  $+69.6344$ & $15.274\pm0.093$ & $15.3143\pm0.0260$ &  6.2 & 2 & d \\
UX Ari                  &  51.64763 &  $+28.7146$ & $19.900\pm0.390$ & $19.7836\pm0.1264$ &  6.3 & 2 & e \\
PSR (*)                &  53.24727 &  $+54.5791$ & $ ~0.611\pm0.013$ & $ ~0.2090\pm0.6583$ & 20.1 & 1 & f \\
HR 1099              &  54.19690 &  $+00.5870$ & $33.880\pm0.470$ & $33.9783\pm0.0349$ &  5.6 & 2 & g \\
V1271 Tau           &  55.95148 &  $+25.0041$ & $ ~7.418\pm0.025$ & $ ~7.4354\pm0.0229$ & 11.4 & 1 & h \\
V 913 Per             &  56.13578 &  $+32.1451$ & $ ~3.119\pm0.104$ & $ ~3.3707\pm0.0986$ & 14.6 & 1 & i \\
V 918 Per             &  56.15392 &  $+32.1126$ & $ ~3.129\pm0.512$ & $ ~3.2215\pm0.0409$ & 12.1 & 2 & i \\
$[$PZH96$] $104 &  56.28318 &  $+32.0671$ & $ ~2.680\pm0.076$ & $ ~2.6052\pm0.0428$ & 14.0 & 2 & i \\
        ...                 & \multicolumn{6}{c}{...}   & \\\hline
\end{tabular}\end{center}
    \small\baselineskip=1.0ex\protect
    (*) --- PSR B0329$+$54;  $G$ is the magnitude copied from the Gaia\,DR3 catalog; $B$ is information about the binary nature, where 1  is a single star, 2  is a binary, 3  is an AGB; $R$  is a reference to a literary source, where
    a~-- Nyu et al. (2011),
    b~-- Asaki et al. (2010),
    c~--  Dhawan et al. (2006),
    d~-- Zhang et al.  (2025),
    e~--  Peterson et al. (2011),
    f~--  Kumar et al. (2025),
    g~-- Lestrade et al. (1999),
    h~-- Melis et al. (2014),
    i~--- Ortiz-Le\'on  et al. (2018).
\end{table}}
%%%%%%
%%%%%%%%%%%%%%%%%%%%%%%%%%%%%%%%%%%%%%%%%%%%%
{
\begin{table}[t]
\caption[]{\small   Data on the proper motions of masers and radio stars. }
\begin{center} \label{t-DATA-mu}
\begin{tabular}{|r|c|c|c|c|c|c|c|c|}\hline
   Star  &  $\mu_\alpha \cos\delta$~(VLBI)   &  $\mu_\delta$~(VLBI)
                 &  $\mu_\alpha \cos\delta$~(Gaia)    &  $\mu_\delta$~(Gaia)  \\
                 &   mas/yr  &  mas/yr  &  mas/yr   &  mas/yr  \\ \hline
SY Scl                   &    $~~5.570\pm0.040$   & $ -7.320\pm0.120$  & $ ~5.719\pm0.137$ & $ -6.740\pm0.101$  \\
S Per                     &    $ -0.490\pm0.350$   & $ -1.190\pm0.330$  &  $ -0.480\pm0.072$ & $ -0.470\pm0.076$ \\
HIP 12469             &    $ -0.150\pm0.006$   & $ -0.264\pm0.006$  & $ -0.423\pm0.011$ &  $ -0.256\pm0.012$ \\
RZ Cas                  &   $ ~2.965\pm0.318$   & $37.543\pm0.185$   & $~2.710\pm0.017$  &  $ 37.476\pm0.024$  \\
UX Ari                   &    $44.960\pm0.130$   & $-102.33\pm0.09$   & $46.794\pm0.153$   &  $ -102.876\pm0.112$  \\
PSR (*)                  &   $16.960\pm0.011$   & $-10.382\pm0.022$  & $~5.020\pm0.582$  &  $-2.503\pm0.657$ \\
HR 1099                &   $ -31.59\pm0.33$    & $-161.69\pm0.31$    & $-32.246\pm0.036$  &  $-162.073\pm0.032$  \\
V1271 Tau             &   $19.860\pm0.050$  & $ -45.41\pm0.16$     & $19.860\pm0.032$   &  $ -45.603\pm0.021$ \\
V 913 Per              &   $~2.458\pm0.047$  & $ -7.270\pm0.133$   & $ ~4.086\pm0.104$  &  $ ~-6.854\pm0.063$  \\
V 918 Per              &   $~4.857\pm0.335$  & $-6.750\pm0.488$    & $ ~5.002\pm0.043$  &  $ ~-6.948\pm0.025$ \\
$[$PZH96$] $104  &   $~2.370\pm0.080$  & $-8.271\pm0.160$    & $ ~0.947\pm0.047$  &  $ ~-9.363\pm0.031$  \\
        ...                 & \multicolumn{3}{c}{...}   & \\\hline
\end{tabular}\end{center}
\end{table}}
 %%%%%

Following the release of Gaia\,EDR3, Lindegren et al. (2021) proposed a method for accounting for the Gaia parallax zero-point correction separately for the results of the five- and six-parameter solutions. Many authors (e.g., Flynn et al. 2022; Wang et al. 2022) have concluded that the method of Lindegren et al. (2021) yields good results. Alternative procedures for accounting for the Gaia parallax zero-point correction have also been proposed, for example, by Groenewegen (2021), Maiz Apell\'aniz (2022), or Ding et al. (2024). Accounting for this correction is currently in demand, for example, for establishing a high-precision distance scale using various variable stars, where calibration by trigonometric parallaxes is required (e.g., Riess et al. 2022; Lengen et al. 2025).

Important attention is given to monitoring the residual rotation of the Gaia system relative to an inertial coordinate system using VLBI measurements of masers and radio stars (e.g., Lindegren 2020; Bobylev 2022; Lunz et al. 2023; Zhang et al. 2024; 2025). Despite the high accuracy of VLBI measurements, comparisons with optical observations sometimes lead to significant discrepancies in both parallaxes and proper motions. This is due to the presence of a large number of binary systems in the sample; asymptotic giant branch stars and simple giants with extended atmospheres with unevenly distributed maser spots. Therefore, Lindegren (2020) developed a special technique for selecting differences with the smallest discrepancies.

Lunz et al. (2023) showed that the rotation of the Gaia\,DR2 system, determined from radio stars with VLBI observations, is $(\omega_x, \omega_y, \omega_z)=(-0.056, -0.113, 0.033) \pm (0.046, 0.058, 0.053)$~mas/yr. From a comparison of the VLBI measurements with the Gaia\,EDR3 catalog data (which had already been corrected for rotation during the Gaia reduction), these authors found $(\omega_x, \omega_y, \omega_z)=(0.022, 0.065, -0.016) \pm (0.024, 0.026, 0.024)$~mas/yr, and concluded that the $\omega_y$ component is significant at the 2.4$\sigma$ level. Zhang et al. (2025) showed that the values of  the residual rotation parameters depend strongly on the stellar magnitude. The estimate obtained by these authors for stars in the interval $0<G<10.^m5$ is $(\omega_x, \omega_y, \omega_z)=(0.021, 0.052, -0.007) \pm (0.018, 0.020, 0.020)$~mas/yr.

The sample of masers and radio stars in Bobylev's (2022) paper consisted of 126 objects. A number of new high-precision VLBI measurements of such objects have now been published, expanding the sample to 151 sources. The goal of this paper is to compile the most complete sample to date of masers and radio stars with VLBI-measured trigonometric parallaxes and proper motions in common with the Gaia\,EDR3 and Gaia\,DR3 catalogs. Conducting a new comparison of VLBI measurements with Gaia data in order to refine the Gaia parallax zero-point correction and the residual rotation of the Gaia system relative to the inertial coordinate system.

\section*{DATA}\label{data}
VLBI observations of masers to measure their trigonometric parallaxes and proper motions are primarily conducted at frequencies of 6.7 and 12.2 GHz, where methanol (CH$_3$OH) masers emit; at 22.2 GHz, where water vapor (H$_2$O) masers emit; and at 43 GHz, where silicon monoxide (SiO) masers emit. In addition to masers, VLBI observations of continuum radio stars, typically conducted at frequencies of 5 and 8 GHz, are of interest. Maser emission is produced by stars surrounded by extensive gas and dust envelopes and extended atmospheres. These can be very young protostars or already formed young T Tauri stars, as well as older OH/IR and AGB stars.

Most important for the purposes of this paper, the VLBI method measures absolute trigonometric parallaxes and proper motions of masers and radio stars, referenced to extragalactic reference sources. The achieved error in determining trigonometric parallaxes averages about 10 mas (e.g., Reid et al. 2019).

The sample analyzed in Bobylev's paper (2022) consisted of 126 masers and radio stars. We note the publications that made it possible to significantly expand the sample. These are VLBI measurements of H$_2$O masers in the envelope of the BX\,Cam variable by Xu et al. (2022); measurements of the trigonometric parallaxes of the pulsars PSR\,B0531+21, AR\,Sco and PSR\,B0329+54 in the papers of Lin et al. (2023), Jiang et al. (2023) and Kumar et al. (2025); measurements of HD\,199178 and AR\,Lac in the paper of Chen et al. (2023); New measurements of the young stellar system $\rho$~Oph S1 by Ord\'o\~nez-Toro et al. (2024), measurements of three AGB stars NSV\,17351, AW\,Tau, and IRC$-$30363 by Nakagawa et al. (2023; 2024), and finally, new measurements of 12 stars from the collection of Zhang et al. (2025) with the corresponding bibliographic references.

The original trigonometric parallax data for the stars in our sample are given in Table~\ref{t-DATA-pi}. Table~\ref{t-DATA-mu} contains the proper motion data for these sources. The first rows are given, and the rest will be available as spreadsheets. Note that VLBI parallax measurements are missing for several sources, while proper motions are available for all stars.

%%%%%%%%%%%%%%%%%%%%%%%%%%%%%%%%%%%%%%%%%%%%% FIG.1:
\begin{figure}[t]{\begin{center}
 \includegraphics[width=0.55\textwidth]{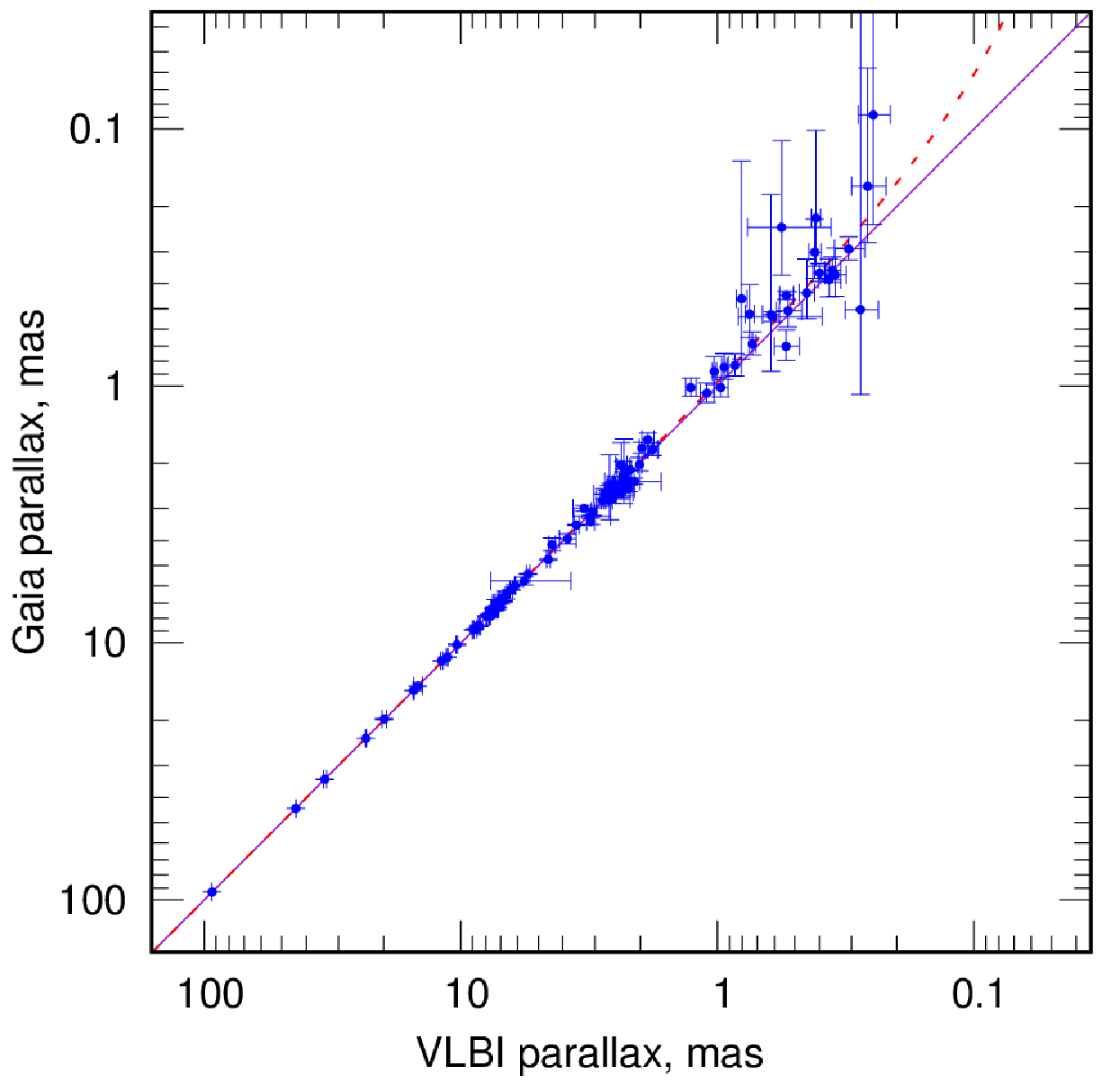}
 \caption{
Parallaxes of radio stars from the Gaia\,DR3 catalogue depending on their VLBI parallaxes, the solid line corresponds to a correlation with a coefficient of 1, the dotted lines correspond to the results of solving equation~(\ref{EQ-1}) using these data (see table \ref{t-pi-1}).
  } \label{f-par}
\end{center}}\end{figure}
%%%%%%%%%%%%
%%%%%%%%%%%%%%%%%%%%%%%%%%%%%%%%%%%%%%%%%%%%% FIG.2:
 \begin{figure}[t]{\begin{center}
  \includegraphics [width=0.45\textwidth] {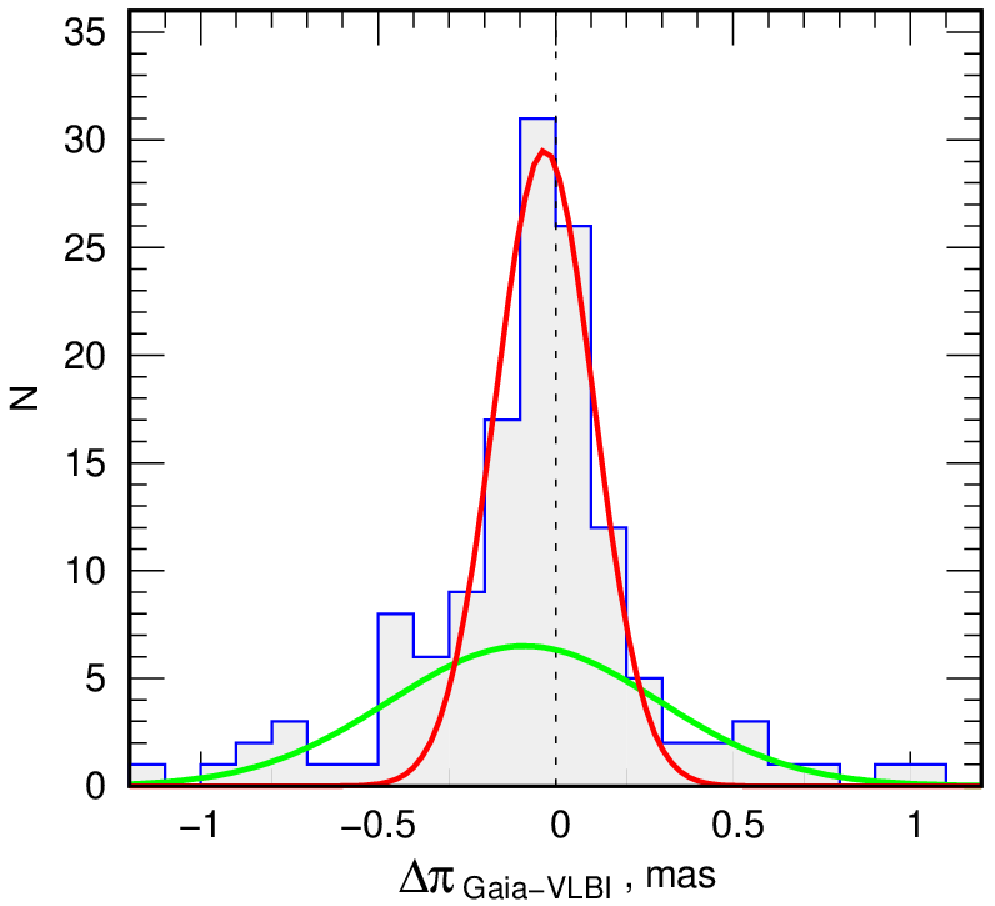}
 \caption{ Parallax differences of stars of the type ``Gaia--VLBI''.} \label{f-delta-pi}
\end{center}}\end{figure}
%%%%%%%%%%%%
%%%%%%%%%%%%%%%%%%%%%%%%%%%%%%%%%%%%%%%%%%%%%
{
\begin{table}[t]
\caption[]{\small  Values of parameters $a$ and $b$ found from the solution of equation~(\ref{EQ-1}). }
\begin{center} \label{t-pi-1}
\begin{tabular}{|r|c|c|c|c|c|c|c|c|}\hline
  $N_\star$ & $a,$~mas & $b$  & restrictions  & $N_{singl} $ & $N_{bin}$ &  $N_{AGB}$ \\\hline
   134 & $-0.052\pm0.017$  & $1.004\pm0.003$   & $|\Delta\pi|<2.00$~mas   & 65 & 45 &  24 \\
   125 & $-0.044\pm0.014$  & $1.003\pm0.002$   & $|\Delta\pi|<0.75$~mas   & 60 & 44 &  21 \\
   104 & $-0.039\pm0.011$  & $1.002\pm0.002$   & $|\Delta\pi|<0.35$~mas   & 50 & 36 & 18 \\       \hline
\end{tabular}\end{center}
\end{table}}
%%%%%%

\section*{RESULTS AND DISCUSSION}\label{results}
\subsection*{Parallax Comparison}\label{PAR}
To find the relationship between the parallaxes of the two systems, we solve the following equations using the least squares method (LSM):
\begin{equation}
\begin{array}{lll}
\pi_{\rm (Gaia)}=a+b\cdot\pi_{\rm (VLBI)}.
\label{EQ-1}
\end{array}
\end{equation}
For this purpose, several subsamples with different constraints on the parallax differences of the Gaia--VLBI type were used.
The least-squares solution was searched for with weights $p$ of the following form: $p=1/(\sigma^2_{\pi \rm (Gaia)}+\sigma^2_{\pi \rm (VLBI)})$, where $\sigma_\pi$ are the corresponding parallax measurement errors. Such a weighting system was proposed by Xu et al. (2019). The results obtained in Bobylev (2022) using various weighting systems showed that using such weights allows one to obtain estimates of the sought-for parameters with the smallest errors.

The parameter estimates obtained from the least-squares solution of equation~(\ref{EQ-1}) $a$ and $b$ are given in Tables~\ref{t-pi-1} and \ref{t-pi-2}, and are also shown in Figs~\ref{f-par} and \ref{f-delta-pi}. Table~\ref{t-pi-1} gives the number of stars $N_\star$ used,  the values $a$ and $b$ found from the solution of equation~(\ref{EQ-1}) for three subsamples. The solutions were obtained in a single iteration, without discarding residuals according to the 3$\sigma$ criterion. In this case, the star HD\,283641 was discarded, its inclusion having a strong negative effect on the results. The solutions obtained in this way are interesting in that the parameter $a$ differs significantly from zero in all three cases. A total of 134 radio stars and masers in our sample have VLBI measurements of their trigonometric parallaxes. Among them, 65 are single $(N_{singl})$, 45 are binaries $(N_{bin})$, and 24 stars belong to the asymptotic giant branch $(N_{AGB})$. The corresponding information for each subsample is given in the last three columns of Table~\ref{t-pi-1}.

Figure ~\ref{f-par} shows the parallaxes of 104 masers and radio stars from the Gaia\,DR3 catalog as functions of their VLBI parallaxes. The dotted line shows the dependence constructed using the data from the third row of Table~\ref{t-pi-1}.

%%%%%%%%%%%%%%%%%%%%%%%%%%%%%%%%%%%%%%%%%%%%%
{
\begin{table}[t]
\caption[]{\small Values of parameters $a$ and $b$ obtained as a result of the least squares solution of equation~(\ref{EQ-1}) in two iterations with rsiduals discarded according to the 3$\sigma$ criterion.
 }
\begin{center} \label{t-pi-2}
\begin{tabular}{|r|c|c|c|c|c|c|c|}\hline
  $N_\star$ &               $a,$~mas &    $b$  &     restrictions  & ${\overline \Delta\pi}\pm\varepsilon_{\Delta\pi},$~mas & $\sigma_{\Delta\pi},$~mas \\\hline
           117 & $-0.070\pm0.018$  & $1.007\pm0.002$   & $|\Delta\pi|<2.00$~mas& $-0.083\pm0.033$ & 0.378 \\
           118 & $-0.051\pm0.018$  & $1.007\pm0.002$   & $|\Delta\pi|<0.75$~mas& $-0.041\pm0.022$ & 0.246  \\
           102 & $-0.038\pm0.011$  & $1.002\pm0.002$   & $|\Delta\pi|<0.35$~mas& $-0.031\pm0.014$ & 0.139 \\
      \hline
\end{tabular}\end{center}\end{table}}
%%%%%%%%%%%%

Table \ref{t-pi-2} presents the results obtained by the least-squares solution of equation~(\ref{EQ-1}) in two iterations with residuals discarded according to the 3$\sigma$ criterion. Stars with relative parallax errors ($\sigma_\pi/\pi$) less than 30\% in each of the catalogs were used. The last columns of the table give  the values of the simple mean $\overline {\Delta\pi}$~masd and the variance of the simple mean $\sigma_{\Delta\pi}$.

Figure~\ref{f-delta-pi} shows a histogram constructed using the parallax differences of radio stars of the type ``Gaia--VLBI''.
This histogram contains two Gaussians, whose parameters are given in the first and third rows of Table~\ref{t-pi-2}. The green line is based on the analysis of 135 stars with a simple mean difference of $\overline {\Delta\pi}=-0.083$~mas and a variance of $\sigma_{\Delta\pi}=0.378$~mas, and the red line is based on the analysis of 102 stars with a mean
$\overline {\Delta\pi}=-0.031$~mas and a variance of $\sigma_{\Delta\pi}=0.139$~mas.

%%%%%%%%%%%%%%%%%%%%%%%%%%%%%%%%%%%%%%%%%%%%%
{
\begin{table}[t]
\caption[]{\small
 Values of parameters $a$ and $b$ obtained as a result of the least squares solution of equation~(\ref{EQ-1}) in two iterations with
residuals discarded according to the 3$\sigma$ criterion under additional constraints.
 }
\begin{center} \label{t-pi-3}
\begin{tabular}{|r|c|c|c|c|c|c|c|}\hline
  $N_\star$ &               $a,$~mas &                       $b$  &                restrictions    \\\hline
          97 & $-0.061\pm0.015$  & $1.006\pm0.001$   & $|\Delta\pi|<2.00$~mas,    $N_{AGB}=0$   \\
          60 & $-0.044\pm0.024$  & $1.006\pm0.003$   & $|\Delta\pi|<2.00$~mas,    $G\leq10.^m5 $ \\
              &&& \\
         96 & $-0.054\pm0.017$  & $1.001\pm0.001$   & $|\Delta\pi|<0.75$~mas,   $N_{AGB}=0$   \\
         62 & $-0.075\pm0.024$  & $1.007\pm0.003$   & $|\Delta\pi|<0.75$~mas,   $G\leq10.^m5 $  \\
             &&& \\
        86 & $-0.037\pm0.012$  & $1.002\pm0.002$   & $|\Delta\pi|<0.35$~mas,   $N_{AGB}=0$   \\
        52 & $-0.049\pm0.015$  & $1.004\pm0.002$   & $|\Delta\pi|<0.35$~mas,   $G\leq10.^m5 $   \\
      \hline
\end{tabular}\end{center}\end{table}}
%%%%%%%%%%%%

Table \ref{t-pi-3} presents the results obtained by the least-squares solution of equation~(\ref{EQ-1}) in two iterations with residual rejection according to the 3$\sigma$ criterion under additional constraints. Here, solutions were obtained a)~without AGB stars and b)~with the rejection of faint stars, i.e., for stars in the interval $G\leq10.^m5$. Such additional constraints were adopted according to the recommendations of Xu et al. (2019) and Zhang et al. (2025). As can be seen from Table~\ref{t-pi-3}, the use of two additional constraints has a favorable effect on the results. For example, the errors in the parameter $a$ decreased when AGB stars were rejected, despite the fact that the number of stars decreased. However, a radical improvement in the solution results was not observed. In general, we can conclude that the best estimate of the Gaia parallax zero-point shift found in this paper is $\Delta\pi=-0.038\pm0.011$~mas, which was obtained with the smallest error (bottom row of the table~\ref{t-pi-2}) using strict constraints.

We note the work of Ding et al. (2025), who obtained estimates of $\Delta\pi$ based on several samples of stars compared with the Gaia\,DR3 catalog. An analysis of 44 binary systems with orbital parallaxes yielded $\Delta\pi=-0.039\pm0.010$~mas. A sample of 31 variable stars with trigonometric parallax measurements taken with the Hubble Space Telescope yielded $\Delta\pi=-0.032\pm0.014$~mas. The estimate of $\Delta\pi=-0.038\pm0.011$~mas in the present study is in very good agreement with these two results.

Ding et al. (2025) also considered the original sample of 102 masers with measured VLBI parallaxes (this sample is described in Xu et al. (2019)). After excluding stars that do not satisfy the imposed constraints ($G<13^m, \sigma_\pi/\pi<20\%, |b|<20^\circ$), 53 bright sources remained, from which $\Delta\pi=-0.015\pm0.011$~mas were found. The significant difference with the estimate of the present work can be explained by the fact that Ding et al. (2025) analyzed a limited (mainly in magnitude) sample of masers. Note that based on 90 masers and radio stars taken from the work of Xu et al. (2019), Bobylev (2022) found $\Delta\pi=-0.022\pm0.017$~mas.

\subsection*{Comparison of Proper Motions}\label{mu}
To determine the three angular velocities of mutual rotation of the two systems around the equatorial coordinate axes $\omega_x,\omega_y,\omega_z$, we use equations of the following form:
\begin{equation}
\begin{array}{lll}
\Delta\mu_\alpha\cos\delta= - \omega_x\cos\alpha\sin\delta - \omega_y\sin\alpha\sin\delta + \omega_z\cos\delta, \\
\Delta\mu_\delta= + \omega_x\sin\alpha - \omega_y\cos\alpha,
\label{Gaia-VLBI}
\end{array} \end{equation}
where the left-hand sides of the equations contain differences of the type ``Gaia--VLBI''. We solve the system of conditional equations (Gaia--VLBI) using the least-squares method with weights of the form $p=1/(\sigma^2_{\mu_{\rm (Gaia)}}+\sigma^2_{\mu_{\rm (VLBI)}})$.

\begin{figure}[t]{\begin{center}
  \includegraphics [width=0.45\textwidth] {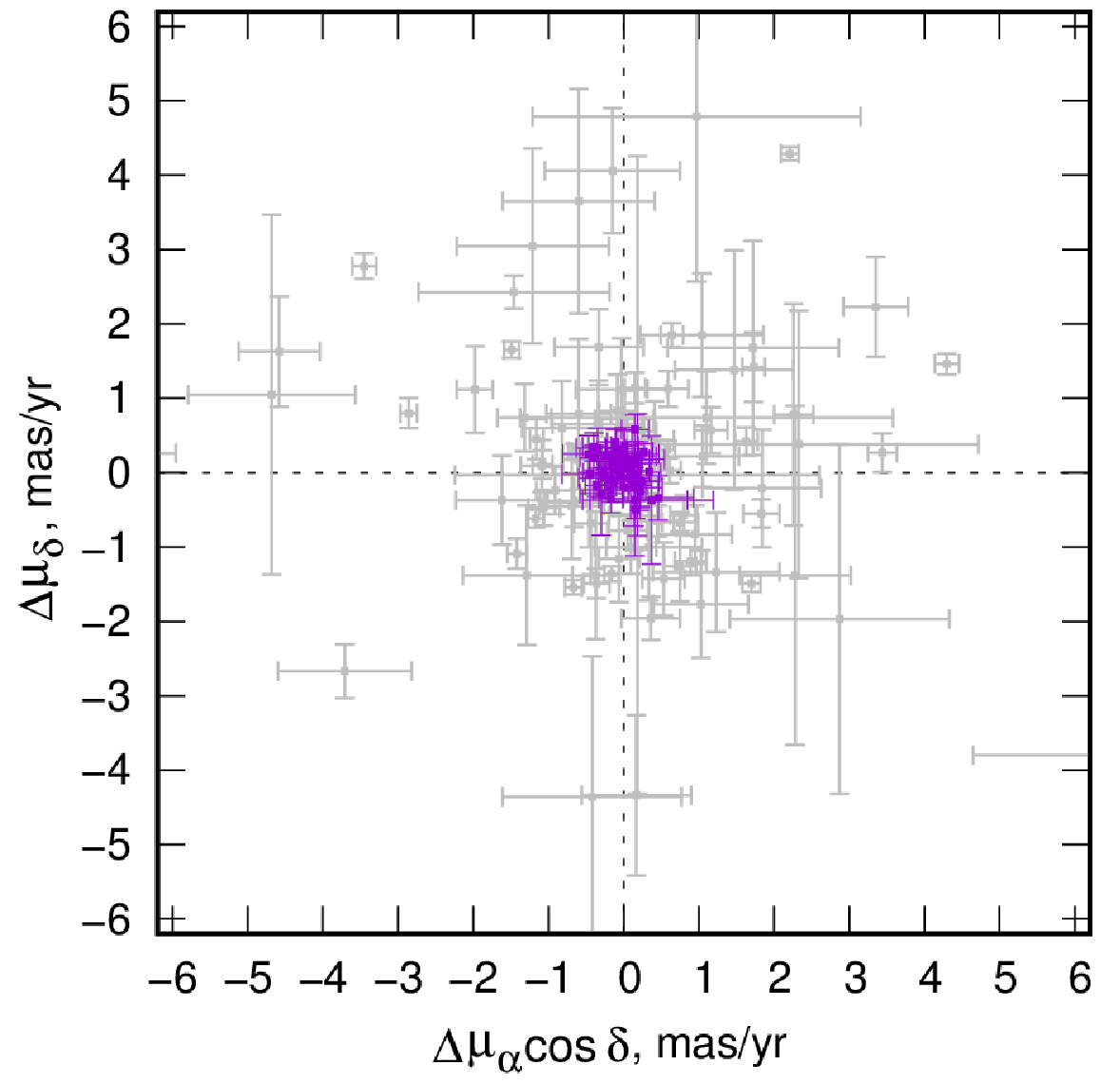}
 \caption{ Differences in the proper motions of stars of the type ``Gaia--VLBI''.} \label{f-Mux-Muy}
\end{center}}\end{figure}

In total, our sample contains proper motion measurements for 150 radio stars. The proper motion differences for stars of the Gaia minus VLBI type are shown in Fig.~\ref{f-Mux-Muy}, where the proper motion differences for stars in the entire sample with their errors are shown in gray. The following values were found from the proper motion differences for 52 stars:
$(\omega_x,\omega_y,\omega_z)=(0.054, 0.056, -0.013)\pm(0.027, 0.034, 0.026)$~mas/yr. In this solution, we used a constraint on the errors in the proper motion differences of $\sqrt {\Delta \mu^2_\alpha \cos\delta+\Delta \mu^2_\delta}\leq0.6$~mas/yr, as well as on the random errors in determining the proper motions, which for each of the components should not exceed 2~mas/yr. These 52 stars are shown in color in the center of Fig.~\ref{f-Mux-Muy}. Note that when searching for the least squares solution using the 3$\sigma$ criterion, two stars, WH\,502 and SY\,Scl, were discarded.

%%%%%%%%%%%%%%%%%%%%%%%%%%%%%%%%%%%%%%%%%%%%% FIG.3:
%%%%%%%%%%%%

Thus, the situation is significantly more dramatic compared to the previous case, where the Gaia parallax zero-point bias was estimated. Here, approximately 65\% of the data had to be discarded to obtain rotation parameter estimates with acceptable errors.

It is worth noting the fundamental agreement between our estimates of $(\omega_x,\omega_y,\omega_z)$ and the results of Lunz et al. (2023) and Zhang et al. (2025), which were cited in the introduction. That is, $\omega_x$ and $\omega_y$ have positive  values of $\sim0.05$~mas/yr, while $\omega_z$~ has a small negative value. However, in our solution, the errors in determining the sought-for parameters are slightly larger than those of Lunz et al. (2023) and Zhang et al. (2025). At the same time, the errors have been significantly reduced compared to the estimates in Bobylev (2022), who used a similar approach to find
$(\omega_x,\omega_y,\omega_z)=(0.06, 0.08, -0.10) \pm (0.06, 0.07, 0.08)$~mas/yr.

\section*{CONCLUSION}\label{discussion}
The most complete list to date of masers and radio stars with VLBI measurements of their trigonomic parallaxes and proper motions, shared with the Gaia EDR3 and Gaia DR3 catalogs, has been compiled. The list contains 151 stars.

A significant expansion of the sample of sources required for analysis allowed us to estimate the systematic bias of the Gaia parallax zero point at a significance level greater than 3$\sigma$. Thus, using the differences in the trigonometric parallaxes of 102 Gaia--VLBI stars, we found that $\Delta\pi=-0.038\pm0.011$~mas, while the parallactic scale factor is close to unity, $b=1.002\pm0.002$.

Using the differences in the proper motions of Gaia--VLBI stars, we obtained estimates of their relative rotation rates
$\omega_{x,y,z}=(0.054, 0.056, -0.013)\pm(0.027, 0.034, 0.026)$~mas/yr. These estimates indicate the absence of any significantly nonzero rotation of the Gaia\,DR3 system relative to the inertial coordinate system, which in this case is represented by the VLBI measurements.

\medskip
The author thanks the reviewer for helpful comments that helped improve the article.

\bigskip\medskip{BIBLIOGRAPHY}\medskip

{\begin{enumerate}\small

\item
M. Andriantsaralaza, S. Ramstedt, W.H.T. Vlemmings, and E. De Beck, Astron. Astrophys. {\bf 667}, A74 (2022).

 \item
Y. Asaki, S. Deguchi, H. Imai, et al., Astrophys. J. {\bf 721}, 267 (2010). %SPer

 \item
V.V. Bobylev, Astron. Lett. {\bf 45}, 10 (2019).

 \item
V.V. Bobylev, Astron. Lett. {\bf 48}, 790 (2022).

\item
V.C. Chan, J. Bovy, Mon. Not. R. Astron. Soc. {\bf 493}, 4367 (2020).

 \item
W. Chen, B. Zhang, J. Zhang, et al., Mon. Not. R. Astron. Soc. {\bf 524}, 5357 (2023).

\item
L.-C. Deng, H.J. Newberg, C. Liu, et al., Res. Astron. Astrophys. {\bf 12}, 735  (2012). % LAMOST

\item
Y. Ding , S. Liao, Q. Wu, Z. Qi, and Z. Tang, Astron. Astrophys. {\bf 691}, A81 (2024).

\item
Y. Ding , S. Liao, S. Wen, and Z. Qi, Astron. J. {\bf 169}, 211 (2025).

\item
V. Dhawan, A. Mioduszewski, and M. Rupen, Proc. VI Microquasar Workshop: Microquasars and
Beyond. September 18--22, 2006, Como, Italy., id.52.1 (2006).

\item
C. Flynn, R. Sekhri, T. Venville, et al., Mon. Not. R. Astron. Soc. {\bf 509}, 4276 (2022).

 \item
Gaia Collab. (T. Prusti, J.H.J. de Bruijne, A.G.A. Brown,  et al.), Astron. Astrophys. {\bf 595}, 1 (2016). %Gaia

 \item
Gaia Collab. (A.G.A. Brown, A. Vallenari, T. Prusti, et al.), Astron. Astrophys. {\bf 616}, 1 (2018). %Gaia DR2

\item
Gaia Collab. (A.G.A. Brown, A. Vallenari, T. Prusti, et al.), Astron. Astrophys. {\bf 649}, 1 (2021). %Gaia EDR3

 \item
Gaia Collab. (A. Vallenari, A.G.A. Brown, T. Prusti, et al.), Astron. Astrophys. {\bf 674}, A1 (2023). %Gaia DR3

\item
D. Graczyk, G. Pietrzy$\acute{\rm n}$ski, W. Gieren, et al., Astrophys. J. {\bf 872}, 85 (2019).

 \item
M.A.T. Groenewegen, Astron. Astrophys. {\bf 619}, A8 (2018).

\item
M.A.T. Groenewegen, Astron. Astrophys. {\bf 654}, A20 (2021).

 \item
Y. Huang , H. Yuan, T.C. Beers, et al., Astrophys. J. {\bf 910}, L5 (2021).

 \item
P. Jiang, L. Cui, J. Yang, et al., Mon. Not. R. Astron. Soc. {\bf 520}, 2942 (2023).

 \item
S. Khan, A. Miglio, B. Mosser, et al., Astron. Astrophys. {\bf 628}, A35 (2019).

 \item
A. Kumar, A.T. Deller, P. Jain, and J. Mold\'on, Publ. Astron. Soc. Australia {\bf 42}, e098 (2025).

\item
A.C. Layden, G.P. Tiede, B. Chaboyer, C. Bunner, and M.T. Smitka, Astron. J. {\bf 158}, 105 (2019).

\item
B. Lengen, R.I. Anderson, M.C. Reyes, and G. Viviani,  arXiv: 2509.16331 (2025).

\item
J.F. Lestrade,  R.A. Preston, D.L. Jones, et al.,  Astron. Astrophys. {\bf 344}, 1014 (1999).

\item
R. Lin, M.H. van Kerkwijk, F. Kirsten, et al., Astrophys. J. {\bf 952}, 161 (2023).

 \item
L. Lindegren, J. Hern\'andez, A. Bombrun, Astron. Astrophys. {\bf 616}, A2 (2018).

\item
L. Lindegren, Astron. Astrophys. {\bf 633}, A1 (2020).

\item
L. Lindegren, U. Bastian, M. Biermann, et al., Astron. Astrophys. {\bf 649}, A4 (2021).

\item
H.W. Leung, J. Bovy, Mon. Not. R. Astron. Soc. {\bf 489}, 2079 (2020).

\item
S. Lunz, J.M. Anderson, M.H. Xu, et al., Astron. Astrophys. {\bf 676}, A11  (2023).

\item
S.R. Majewski, R.P. Schiavon, P.M. Frinchaboy, et al., Astron. J. {\bf 154}, 94 (2017). %APOGEE

\item
Maiz Apell\'aniz, Astron. Astrophys. {\bf 657}, A130  (2022).

\item
C. Melis, M.J. Reid, A.J.Mioduszewski, J.R. Stauffer, and  G.C. Bower, Science {\bf 345}, 1029 (2014).

\item
T. Muraveva, H.E. Delgado, G. Clementini, et al., Mon. Not. R. Astron. Soc. {\bf 481}, 1195 (2018).

\item
A. Nakagawa, A. Morita, N. Sakai, et al., PASJ {\bf 75}, 529 (2023).

\item
A. Nakagawa, T. Kurayama, H. Sudou, Hiroshi,  and G. Orosz, Cosmic Masers: Proper Motion Toward the Next-Generation Large Projects, held 20--24 March, 2023 in Kagoshima, Japan. Proc. IAU, {\bf 380}, 300 (2024).

\item
D. Nyu, A. Nakagawa, M. Matsui, et al., PASJ {\bf  63}, 63 (2011).

\item
J. Ord\'o\~nez-Toro, S.A. Dzib, L. Loinard, et al., Astron. J. {\bf 167}, 108 (2024).

\item
G.N. Ortiz-Le\'on,  L. Loinard,  S.A. Dzib, et al., Astrophys. J. {\bf 865}, 73 (2018).

\item
W.M. Peterson, R.L. Mutel,  J.F. Lestrade,  M. G\:udel, and W.M. Goss, Astrophys. J.  {\bf 737}, 104 (2011).

 \item
M.H. Pinsonneault, Y.P. Elsworth, J. Tayar, et al., Astrophys. J. Suppl. Ser. {\bf 239}, 32 (2018).

 \item
F. Ren, X. Chen, H. Zhang, et al., Astrophys. J. {\bf 911}, L20 (2022).

 \item
M.J. Reid, K.M. Menten, A. Brunthaler, et al., Astrophys. J. {\bf 885}, 131 (2019).

 \item
A.G. Riess, S. Casertano, W. Yuan, et al., Astrophys. J. {\bf 861}, 126 (2018).

 \item
A.G. Riess, L. Breuval, W. Yuan, et al., Astrophys. J. {\bf 938}, 36 (2022).

 \item
K.G. Stassun, G. Torres, Astrophys. J. {\bf 862}, 61 (2018).

 \item
K.G. Stassun, G. Torres, Astrophys. J. {\bf 907}, L33 (2021).

 \item
C. Wang, H. Yuan, and Y. Huang, Astron. J. {\bf 163}, 149 (2022).

  \item
S. Xu, B. Zhang, M.J. Reid, X. Zheng, and G. Wang, Astrophys. J. {\bf 875}, 114 (2019).

  \item
S. Xu, H. Imai, Y. Yun, et al., Astrophys. J. {\bf 941}, 105 (2022).

 \item
J. Zhang, B. Zhang, S. Xu, et al., Mon. Not. R. Astron. Soc.  {\bf 529}, 2062 (2024).

 \item
J. Zhang, B. Zhang, S. Xu, and X. Mai, arXiv: 2506.18758 (2025).

 \item
J.C. Zinn, M.H. Pinsonneault, D. Huber, and D. Stello, Astrophys. J. {\bf 878}, 136 (2019).

\end{enumerate} }
\end{document}